\documentclass[aps,prl,reprint,amsmath,amssymb,showpacs,superscriptaddress]{revtex4-1}

\usepackage{graphicx}
\usepackage{dcolumn}
\usepackage{bm}
\usepackage{color}
\usepackage{ulem}

\begin{document}

\title{Rheology of ring polymer melts: From linear contaminants to ring/linear blends}

\author{Jonathan D. Halverson}
\affiliation{Max Planck Institute for Polymer Research, Ackermannweg 10, 55128 Mainz, Germany}

\author{Gary S. Grest}
\affiliation{Sandia National Laboratories, Albuquerque, NM 87185, USA}

\author{Alexander Y. Grosberg}
\affiliation{Department of Physics, New York University, 4 Washington Place, New York, NY 10003, USA}

\author{Kurt Kremer}
\altaffiliation{The present manuscript has been accepted by \textit{Physical Review Letters}. After it is published, it will be found at \url{http://prl.aps.org}. Corresponding author e-mail: kremer@mpip-mainz.mpg.de}
\affiliation{Max Planck Institute for Polymer Research, Ackermannweg 10, 55128 Mainz, Germany}

\pacs{83.80.Tc, 83.10.Rs, 83.80.Sg}


\begin{abstract}
Ring polymers remain a major challenge to our current understanding of polymer dynamics.
Experimental results are difficult to interpret because of the uncertainty in the purity and dispersity of the sample.
Using both equilibrium and non-equilibrium molecular dynamics simulations we have systematically
investigated the structure, dynamics and rheology of
perfectly controlled ring/linear polymer blends with chains of such length and flexibility that the number of entanglements is up to about 14 per chain, which is comparable to experimental systems examined in the literature.
The smallest concentration at
which linear contaminants increase the zero-shear viscosity of a ring polymer melt of these chain lengths by 10\% is approximately one-fifth of their overlap concentration.
When the two architectures are present in equal amounts the viscosity of the blend is approximately twice as large as  that of the pure linear melt. At this concentration
the diffusion coefficient of the rings is found to decrease dramatically, while
the static and dynamic properties of the linear polymers are mostly unaffected. Our results are
supported by a primitive path analysis.
\end{abstract}

\maketitle

While much has been learned about linear
and branched polymers~\cite{deGennes79,Doi86}, a comparable understanding of ring
or cyclic polymers is lacking. Ring polymers, as they do not have free ends, represent the simplest model system
where reptation is completely suppressed. Also, mitochondrial and plasmid
DNA are usually cyclic, and melts of rings are considered highly relevant model
systems to understand chromatin folding in the cell
nucleus~\cite{Chromosome_territories_Cremer,Rosa_Everaers_PLOS_2008}. This
makes ring polymers perfect test cases for both fundamental polymer
and bio physics.

Early experimental studies on pure ring polymer melts gave inconsistent
results most likely because the samples were contaminated with
linear chains~\cite{Hild83,Roovers83,Hild87}. Also the existence of self-knots
could not be controlled. More recently experiments have been conducted~\cite{Rubinstein_Nature_2008}
based on new characterization and purification techniques~\cite{Pasch,Chang00}.
For melts of nonconcatenated polystyrene rings with
molecular weight (MW) to entanglement MW ratios of
9.2 and 11.3, where the entanglement MW is 17500 g/mol, Kapnistos et al.~\cite{Rubinstein_Nature_2008} reported that the
stress relaxation modulus, $G(t)$, follows a power-law decay with no
sign of a rubbery plateau. The authors used scaling arguments to show $G(t) \sim t^{-2/5}$, a result in agreement with
the data up to the terminal time.
Milner and Newhall~\cite{Milner10} introduced the ``diffusion of centrality'' concept and mapped the ring conformations to annealed tree-like structures and found a similar prediction of $G(t) \sim t^{-1/2}$.
Kapnistos et al.~\cite{Rubinstein_Nature_2008} also reported that
the smallest concentration of linear contaminants that affects the rheology of
the ring melt is almost two decades below the overlap concentration of the linear chains.
Despite the synthetic effort, the characterization and control of the experimental systems including polydispersity, knotting, concatenation and linear contaminants is far from perfect. Because of this, computer simulations of optimized models, which by now easily reach effective experimental molecular weights, are perfect to test concepts for precisely defined systems under well-controlled conditions.
Our own recent simulations~\cite{halverson_part2} of a melt of nonconcatenated and unknotted ring polymers have shown
that $G(t) \sim t^{-\alpha}$ with $\alpha$ decreasing from $0.5$ to $0.45$ with increasing chain length.

Here we employ molecular dynamics (MD) simulations to study the structure, dynamics and rheology of
ring/linear polymer blends of equal chain length. We consider two lengths of $N=200$ and 400 monomers
per chain. For the model used here the entanglement length of a melt of
linear polymers is $N_e=28 \pm 1$~\cite{Everaers04} which corresponds to $N/N_e \approx 7.1$ and
14.3 entanglements per chain. For this a bond bending potential ($k_{\theta}=1.5~\epsilon$) along the chains is introduced, leading to a Kuhn length of $l_k\cong 2.79~\sigma$~\cite{Everaers04}, $\sigma$ being the unit of length. $N_e$  is determined by a primitive path analysis~\cite{Everaers04,sukumaran_ppa}, which is known to yield $N_e$ values which properly reproduce rheological data~\cite{Hoy09,Everaers11}. Our systems are perfectly monodisperse, unknotted and nonconcatenated, allowing for a rather
stringent test of currently discussed concepts. Previous simulations of such mixtures have only considered short chain lengths
and did not measure any rheological properties~\cite{Iyer07,subramanian2008b}. While different polymer melts can be related to each other by the $N/N_e$ ratio, we note that for the present comparison to experiment~\cite{Rubinstein_Nature_2008} also the ratio of the Kuhn length and the packing length $l_k/p$ are not that different, namely 6.5 for our simulation model~\cite{Everaers04} and 3.8 for a polystyrene melt~\cite{fetters99}.

\begin{figure}[]
\includegraphics[scale=1.0]{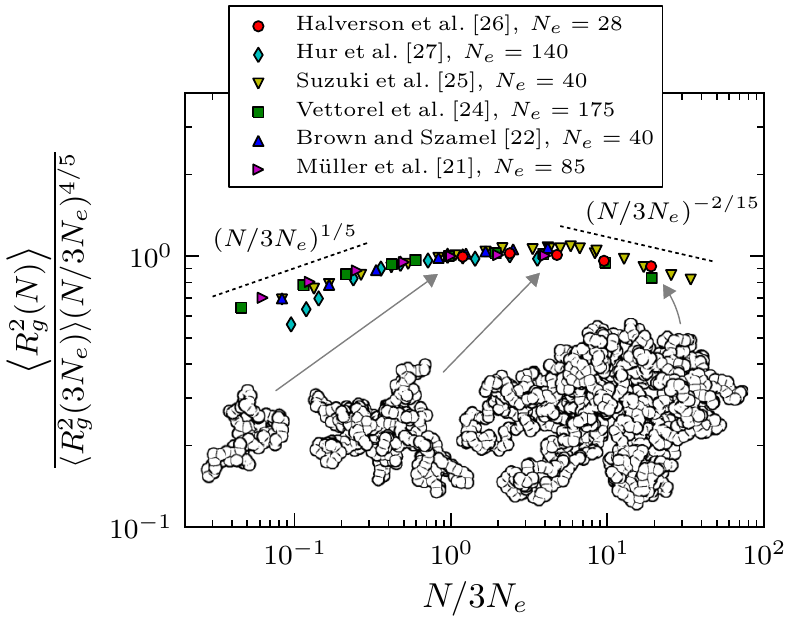}
\caption{Universal behavior of $\left\langle R_g^2(N) \right\rangle$ for
pure ring polymer melts. The data were obtained using different simulation methods
and different models. The reference line with slope $1/5$ corresponds to the Gaussian
regime while that with a slope of $-2/15$ corresponds to the collapsed regime.
Representative conformations from Ref.~\cite{halverson_part1} are shown.}
\label{universal_3Ne}
\end{figure}

The topological constraint that a ring must remain unknotted
and nonconcatenated leads to nontrivial behavior even for the
static properties of a melt or concentrated solution of rings. Rings
are found to be approximately Gaussian at short chain lengths, while for larger lengths the
nonconcatenation dominates the conformational statistics. Cates and Deutsch~\cite{Cates_Deutsch_1986} conjectured
that the exponent in the mean-square gyration radius,
$\left\langle R_g^2 \right\rangle \sim N^{2\nu}$, should be less than $\nu=1/2$ and greater than 1/3 and used a simple free energy argument to arrive at a value of 2/5, which
was later supported by simulation~\cite{Muller96,Brown98b} and
experiment~\cite{arrighi2004} for systems with less than 13 entanglements per chain.
However, for larger rings a scaling of
$\left\langle R_g^2 \right\rangle \sim N^{2/3}$ has been shown~\cite{Vettorel09b,Suzuki09,halverson_part1}.
Altogether we expect a smooth crossover from a Gaussian regime ($\nu=1/2$) via
a regime with $\nu = 2/5$ for rings of length of
a few $N_e$ to the ``crumpled globule'' regime ($\nu = 1/3$) for rings significantly exceeding $N_e$.
The universal scaling behavior of $\left\langle R_g^2(N) \right\rangle$ for a pure ring polymer
melt is demonstrated in Fig.~\ref{universal_3Ne} using results from many different simulations.
Only short-chain atomistic data for polyethylene~\cite{Yoon2010} deviate from the curve~\footnote{The
data of Hur et al.~\cite{Yoon2010} does not fall on the curve in this regime
because polyethylene conformations
are too extended as also known from linear chains.}.
From $N/N_e \approx 15$ the onset of the collapsed regime is clearly
observed in agreement with the predictions of Vettorel et al.~\citep{Vettorel09b}.

We present new MD simulations using the same semiflexible bead-spring model~\cite{Kurt90} as in our previous
work~\cite{halverson_part1,halverson_part2}. The length, time and energy scales are $\sigma$, $\tau$ and $\epsilon$, respectively.
The production runs were carried out using LAMMPS~\cite{plimpton} with a time step of $0.01~\tau$ and an overall monomer density of $\rho=0.85/\sigma^3$. The
largest simulations ran in parallel on 2048 Blue Gene/P cores.
Systems studied range from
$\phi_{\mathrm{linear}} \equiv M_{\mathrm{linear}} / (M_{\mathrm{linear}} + M_{\mathrm{rings}})=0$ to 1,
where $M$ is the number of chains of a given architecture. For $N=200$ the total number
of chains ranged from 200--260 while for $N=400$ the systems were composed of 200--400 chains.
The initial configuration for each blend system with $\phi_{\mathrm{linear}} \leq 0.115$ was created by adding linear chains
at random locations within an equilibrated ring melt configuration. Chains which most closely matched a Gaussian chain were taken from an equilibrated pure linear melt.
For the cases with $\phi_{\mathrm{linear}} \approx 0.25$ and $0.5$ the appropriate number of rings were randomly removed
while for the case with $M_{\mathrm{rings}}=10$ and $M_{\mathrm{linear}}=250$, rings were taken
from an equilibrated pure ring melt and inserted into a linear melt making
sure that the nonconcatenation constraint was observed.
Because these insertions lead to monomers being very nearly overlapping, a short simulation was
carried out for $100~\tau$ while limiting the bead displacement at every step to $0.001~\sigma$. During this short run the box size was increased linearly so as to give the correct density at the final step.
This procedure produces non-equilibrated starting configurations.
Long MD simulations of $4-8\times10^7~\tau$ were performed to equilibrate each system where each architecture
moved at least twice its root-mean-square gyration radius and in some cases more than 20 times this value.

\begin{figure}[]
\includegraphics[scale=1.0]{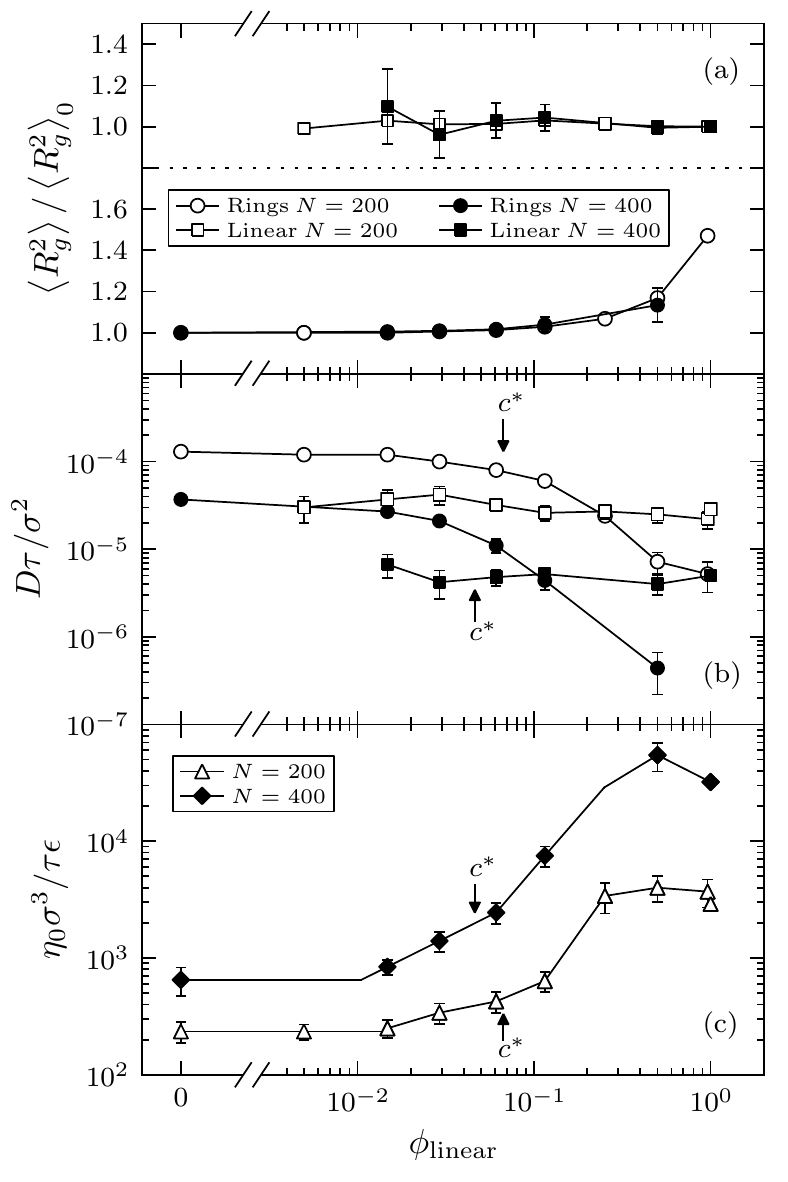}
\caption{(a) Mean-square gyration radii, (b) diffusion coefficients and (c) zero-shear viscosity versus $\phi_{\mathrm{linear}}$. The
overlap concentration of linear chains, $c^{\ast}$, is indicated for the two values of $N$. For the rings with $N=200$ and 400, $\left\langle R_g^2 \right\rangle_0=30.8$ and $52.9~\sigma^2$, respectively, while for the linear chains $\left\langle R_g^2 \right\rangle_0=88.9$ and $180.8~\sigma^2$. Note that the horizontal axis is interrupted. Lines are drawn as a guide for the eye.}
\label{stack}
\end{figure}

Results for the mean-square gyration radius for the rings
and linear chains normalized by their respective pure melt values are shown in Fig. \ref{stack}(a).
For the rings with $N=200$, $\left\langle R_g^2 \right\rangle$ is
found to increase with increasing linear concentration. At $\phi_{\mathrm{linear}} \approx 0.96$, $\left\langle R_g^2 \right\rangle=45.3 \pm 2.2~\sigma^2$, which
is 1.5 times larger than the value of the pure ring melt. For a Gaussian ring
$\left\langle R_g^2 \right\rangle=Nl_kl/12=45.2~\sigma^2$, where $l$ is the average bond length.
For the rings with $\phi_{\mathrm{linear}} \approx 0.96$ the
static structure function scales as $S(q) \sim q^{-2}$ for $2\pi/\langle R_g^2\rangle^{1/2}< q < 2\pi/l_k$, even though the rings cannot sample the whole conformational
space of a Gaussian ring~\cite{RingsNotGaussian}.
For the $N=400$ systems a similar swelling behavior is found for the rings.
The linear chains are found to be Gaussian for all combinations of $N$
and $\phi_{\mathrm{linear}}$. At small values of $\phi_{\mathrm{linear}}$ the rings
are partially collapsed as discussed above. As the
fraction of linear chains increases,
the size of the rings grows because it is entropically favorable for the
linear chains to thread the rings.
At infinite dilution
the nonconcatenation constraint vanishes and
the rings are found to be multiply-threaded and nearly Gaussian~\cite{RingsNotGaussian,Iyer07,yang2010}.

The diffusion coefficients, $D$, which are determined by the
long-time behavior of the mean-square displacement of the center-of-mass of the chains, are shown in Fig. \ref{stack}(b).
The diffusivity of the rings for both values of $N$ is found to steadily decrease with increasing
fraction of linear chains until a dramatic decrease is observed.
With the overlap concentration
of linear chains being $c^{\ast}=\phi_{\mathrm{linear}}^{\ast}\rho=N/(4/3)\pi \left\langle R_g^2 \right\rangle^{3/2}$,
this transition corresponds to approximately $0.1\rho=1.5c^{\ast}$ for $N=200$ and $0.04\rho=0.9c^{\ast}$ for $N=400$.
For $N=400$ the diffusion coefficient of the rings at $\phi_{\mathrm{linear}}=0.5$ is
reduced by a factor of about 75 compared to the pure ring melt. While the linear chains clearly restrict the motion of
the rings, the motion of the linear chains for both values of $N$ is largely independent of
the blend composition, which is consistent with early experimental results~\cite{Tead92}.

Linear chains have free ends and undergo reptation independently of whether the surrounding chains are
rings or linear,
and accordingly their diffusion
is found to be independent of $\phi_{\mathrm{linear}}$.
Rings in a pure melt do not reptate like linear chains. As linear chains
are added to the ring melt, the rings become threaded and the nature of their motion changes. A threaded ring can only diffuse through the release of threads. For a one-thread situation
Mills et al.~\cite{mills1987} have shown that the
diffusion coefficient of the ring is $D \sim N_{\mathrm{ring}}^{-1}N_{\mathrm{linear}}^{-1}$.
At high fractions of linear chains the rings become multiply-threaded and
their diffusion is severely hindered. In this regime the motion of a ring monomer
is coupled to the motion of surrounding linear chains.
This implies
Rouse dynamics for the ring with a monomer relaxation time governed by the reptation relaxation of the linear chains, leading to a relaxation time scaling of $N_{\mathrm{ring}}^2N_{\mathrm{linear}}^3$.
This argument is due to Graessley~\cite{graessley1982} and predicts
$D \sim N_{\mathrm{ring}}^{-1}N_{\mathrm{linear}}^{-3}$.

\begin{figure}[]
\includegraphics[scale=1.0]{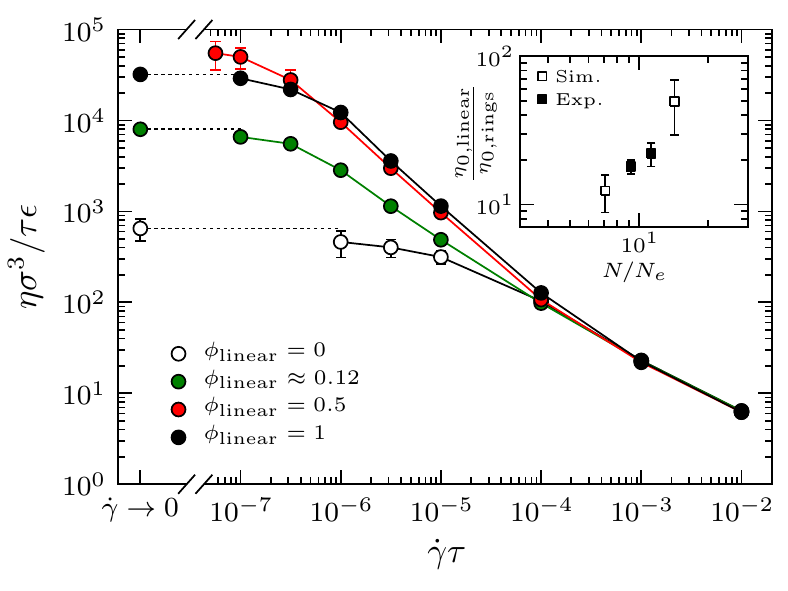}
\caption{Viscosity versus shear rate, $\dot\gamma$, for $N=400$ obtained from non-equilibrium MD simulations~\cite{tuckerman97}.
Zero-shear viscosities obtained from the equilibrium simulations are shown on the far left.
Note that the horizontal scale is interrupted. Inset: Ratio of pure linear to pure ring melt viscosity versus number of entanglements per chain for the simulated systems and the experimental data of Ref.~\cite{Rubinstein_Nature_2008,Rubinstein_privcom_11}.}
\label{nemd}
\end{figure}

The zero-shear viscosity computed
as $\eta_0 =\int_0^{\infty}G(t)dt$ is shown as
a function of $\phi_{\mathrm{linear}}$ in Fig.~\ref{stack}(c).
A striking result is the clear indication that the smallest concentration at which linear
contaminants alter the viscosity of a ring melt considerably (about 10\%) for the chain lengths considered here is
$\phi_{\mathrm{linear}} \approx 1/100$ or $c^{\ast}/5$ with a strong increase around $c^{\ast}$. This
threshold concentration is roughly consistent with the change in $D$ for the rings.
We have confirmed our values of $\eta_0$ by conducting non-equilibrium MD simulations~\cite{tuckerman97}
where simple steady shear is imposed. For these simulations a Nos\'e-Hoover thermostat~\cite{tuckerman97,plimpton} with a relaxation time of $10~\tau$ was used. Note that the thermal velocity is much larger than the largest velocity difference imposed by the shear.
As shown in Fig.~\ref{nemd} for $N=400$, when $\eta(\dot\gamma)$ is extrapolated to $\dot\gamma \rightarrow 0$ the agreement
with $\eta_0$ is very good~\footnote{In Fig. ~\ref{nemd}, $\eta_0$ is not given for the $\phi_{\mathrm{linear}}=0.5$
system because $D$ for the rings is very small and an extremely long
simulation time would be required to resolve $G(t)$.}. Similar agreement is found for $N=200$.
For both values of $N$ the viscosity at $\phi_{\mathrm{linear}} = 0.5$ is larger than the
viscosity at all other concentrations investigated.
For the simulated blends with 14.3 entanglements per chain we
find $\eta(\phi_{\mathrm{linear}}=0.5)/\eta_0(\phi_{\mathrm{linear}} = 1) \gtrapprox 1.8$,
where $\eta$ of the blend is taken from
the non-equilibrium MD simulations (cf. Fig.~\ref{nemd}) which gives a value that is still increasing slightly.
These findings are in good agreement with the experimental results of Roovers~\cite{roovers1988} who showed for
ring/linear blends of polybutadiene with approximately 15.3
entanglements per chain that the maximum in $\eta_0$
occurs at $\phi_{\mathrm{linear}}=0.6$ and $\eta_0(\phi_{\mathrm{linear}} \approx 0.5)/\eta_0(\phi_{\mathrm{linear}} = 1) \approx 2.2$.
The viscosity results in Fig.~\ref{stack}(c) provide a direct macroscopic indication
of the concentration of linear contaminants in experimental samples.
As pointed out by Kapnistos et al.~\cite{Rubinstein_Nature_2008}, the
data also suggest how the viscosity of a linear melt may be tuned
by adding ring polymers.

\begin{figure}[]
\includegraphics[width=8cm]{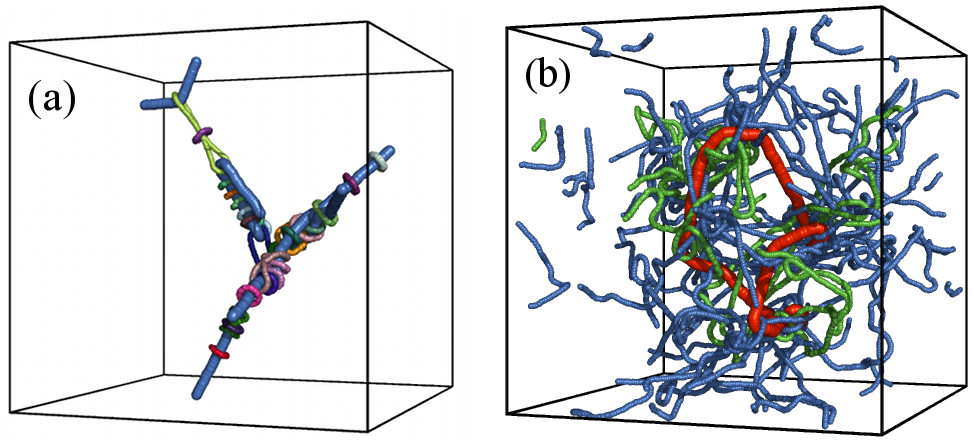}
\caption{
Final configurations from a primitive path analysis for $N=400$.
(a) $\phi_{\mathrm{linear}}=3/203 \approx 0.015$
and all three linear chains (blue) are shown as well as only the rings which did not collapse to points. At
this low concentration of linear chains on average 86\% of the rings are found to be unthreaded.
(b) $\phi_{\mathrm{linear}}=113/226=0.5$ and
a selected ring (red) is shown along with the rings (green) and linear chains (blue) which
it is either threaded by or entangled with. For clarity all other chains are not shown.}
\label{ppa}
\end{figure}

To quantify the extent of threading,
a primitive path analysis~\cite{Everaers04,sukumaran_ppa}
was conducted where the end monomers of the linear chains were fixed and the
rings were allowed to relax freely. This procedure causes the linear chains to
be pulled taut while the rings shrink towards their center with unthreaded rings collapsing to points.
The time scale for the primitive path procedure is $10^3~\tau$ which satisfies the condition of being equal to or faster than $\tau_e = 3200~\tau$, the Rouse time of a linear chain of $N_e$.
Averaging over 10--20 configurations
incremented by $2\times 10^6~\tau$, with $N=200$ the percentage of unthreaded rings
for $\phi_{\mathrm{linear}} \approx 0.03,~0.12,~0.25,~0.96$ is $80.0,~30.3,~11.7,~0.0\%$, respectively.
For $N=400$ with $\phi_{\mathrm{linear}} \approx 0.015,~0.03,~0.12,~0.5$ we find $86.0,~59.0,~7.0,~0.0\%$, respectively.
Fig.~\ref{ppa}(a) shows a final configuration for $\phi_{\mathrm{linear}} \approx 0.015$ where the vast majority of rings are found to be unthreaded.
The sensitivity of a ring melt to linear contaminants is demonstrated by the fact that
the viscosity of this system is already 1.4 times
larger than the pure ring melt value. Fig.~\ref{ppa}(b) shows a final configuration for
$\phi_{\mathrm{linear}} = 0.5$ where a selected ring and the polymers it is entangled with are shown.
Given the large number of entanglements at this composition, the dramatic decrease in the diffusivity of the rings and
the increase in the blend viscosity in comparison to the pure ring melt value
are easily understood.

The present work provides a complete scan of compositions of two different ring polymer/linear polymer melts for dynamical quantities such as viscosity and chain diffusion.
One striking result is that the linear contaminants start significantly affecting the ring melt viscosity at a concentration well below their overlap concentration. This simulation result is in perfect qualitative agreement with the experimental observation of Ref.~\cite{Rubinstein_Nature_2008}: according to both simulation and experiment, there is clearly an effect below the overlap concentration.
However, quantitatively we detect the onset of a viscosity change ($10\%$ increase for rings and linear chains with $N/N_e \approx 10$) at $\phi_{\mathrm{linear}}\approx 0.01$, while Kapnistos et al.~\cite{Rubinstein_Nature_2008} reported a 2-fold viscosity increase in comparison to the ``pure as currently possible rings'' at a much smaller concentration of $\phi_{\mathrm{linear}}=0.0007$. To provide an intuitive picture of these concentrations one can estimate what would be the typical distances between chains. For $\phi_{\mathrm{linear}}=0.0007$ the typical distance between linear chains $(\rho/N)^{-1/3}$ would be about $66~\sigma$ for $N=200$ and $83~\sigma$ for $N=400$. The diameter ($2\left\langle R_g^2 \right\rangle^{1/2}$) of the rings is about $11~\sigma$ and $15~\sigma$ and of the linear chains about $19~\sigma$ and $27~\sigma$, respectively.
Thus two linear chains would be separated on average by about 4--5 ring diameters for $N=200$, or by 4 ring diameters for $N=400$. And, importantly, these rings would not be entangled since they are nonconcatenated and have no free ends. At $\phi_{\mathrm{linear}} = 0.01$, where our data indicate a $10\%$ viscosity increase, distances and chain extensions are all rather similar.

While the two works differ with respect to the onset concentration, fair agreement is
found for the ratio of the pure linear melt viscosity to that of the (almost) pure ring melt as shown in the inset of Fig.~\ref{nemd}~\cite{Rubinstein_Nature_2008,Rubinstein_privcom_11}.
Additionally, the simulation and experimental results for the dependence of $\eta_{0,\mathrm{linear}}/\eta_{0,\mathrm{rings}}$ on $N/N_e$ are consistent not only with one another, but also with the theoretical framework of Ref. ~\cite{Rubinstein_Nature_2008} and our previous result~\cite{halverson_part2} which suggest a power law dependence with power close to 2.

This work significantly benefited from discussions with R. Everaers and M. Rubinstein.
We also thank M. Rubinstein for additional data~\cite{Rubinstein_privcom_11}
concerning Ref.~\cite{Rubinstein_Nature_2008}. KK acknowledges the hospitality of the Center for Soft Matter Research at NYU where part of this work was done.
Funding was provided in part by the Multiscale Materials Modeling (MMM) initiative of the Max Planck
Society. This work is supported by the Laboratory Directed Research and Development program at Sandia
National Laboratories.
Sandia National Laboratories is a multi-program laboratory managed and operated by Sandia Corporation, a wholly owned subsidiary of Lockheed Martin Corporation, for the U.S. Department of Energy's National Nuclear Security Administration under contract DE-AC04-94AL85000.

\bibliography{ring}
\end{document}